\documentclass[iop]{emulateapj}
\usepackage{latexsym}
\usepackage{dcolumn}
\usepackage{bm}
\usepackage{amsmath}
\usepackage{natbib}
\input{epsf}
\newcommand{\be}{\begin{equation}}
\newcommand{\ee}{\end{equation}}
\newcommand{\ba}{\begin{eqnarray}}
\newcommand{\ea}{\end{eqnarray}}
\newcommand{\bi}{\begin{itemize}}
\newcommand{\ei}{\end{itemize}}
\newcommand{\bfi}{\begin{figure}[t]
\epsfxsize=9cm
 \epsffile}
\newcommand{\bfig}{\begin{figure*}[t]
\epsfxsize=15cm
\epsffile}
\newcommand{\efi}{\end{figure}}
\newcommand{\efig}{\end{figure*}}
\newcommand{\no}{\nonumber}

\newcommand{\mpch}{{\rm Mpc}/h}
\newcommand{\hmpc}{h/{\rm Mpc}}
\begin{document}
\title{Accurate Determination of halo velocity bias in simulations and
its cosmological implications}
\author{Junde Chen\altaffilmark{1,4,*}, Pengjie Zhang\altaffilmark{1,2,3,4,$\dagger$},
 Yi Zheng\altaffilmark{5,6}, Yu Yu\altaffilmark{1,4}, Yipeng Jing\altaffilmark{1,2,3,4}}
\altaffiltext{1}{Department of Astronomy, School of Physics and Astronomy, Shanghai
Jiao Tong University, Shanghai, 200240}
\altaffiltext{2}{IFSA Collaborative Innovation Center, Shanghai Jiao Tong
University, Shanghai 200240, China}
\altaffiltext{3}{Tsung-Dao Lee Institute, Shanghai 200240, China}
\altaffiltext{4}{Shanghai Key Laboratory for Particle Physics and
  Cosmology}
\altaffiltext{5}{School of Physics, Korea Institute for Advanced
  Study, Hoegiro 85, Seoul 02455, Korea}
\altaffiltext{6}{Korea Astronomy and Space Science Institute, 776,Daedeokdae-ro, Yuseong-gu, Daejeon 34055, Republic of Korea}
\altaffiltext{*}{jundechen@sjtu.edu.cn}
\altaffiltext{$\dagger$}{zhangpj@sjtu.edu.cn}

\begin{abstract}
A long-standing issue in peculiar velocity
cosmology is whether the  halo/galaxy velocity bias $b_v=1$ at large
scale. The  resolution of this important issue must resort to high precision cosmological
simulations.  However, this is hampered by another long-standing 
``sampling artifact'' problem in volume weighted velocity
measurement.  We circumvent this problem with a hybrid
approach. We first measure statistics free of sampling artifact,  then
link them to  volume weighted statistics in theory, finally solve
for the velocity bias. $b_v$ determined by our method is not only free of
sampling artifact, but also free of  cosmic variance. We apply this method to a
$\Lambda$CDM N-body simulation of  $3072^3$ particles and
$1200 \mpch$ box size. For the first time, we determine the halo
velocity bias to $0.1\%$-$1\%$
accuracy.  Our major findings are as follows: (1) 
$b_v\neq 1$  at $k>0.1\hmpc$. The deviation from unity ($|b_v-1|$) increases with 
$k$. Depending on halo mass and redshift, it may reach $\mathcal{O}(0.01)$ at  $k=0.2\hmpc$ and
$\mathcal{O}(0.05)$ at $k\sim 0.3\hmpc$.  The discovered $b_v\neq
1$ has statistically significant impact on structure growth rate
measurement by  spectroscopic redshift surveys, including DESI, Euclid
and SKA.  (2) Both the sign and the amplitude
of $b_v-1$ depend on mass and  redshift. These results disagree with the peak model prediction
in that $b_v$ has much weaker deviation from unity, varies with
redshift, and can be bigger than unity.  (3) Most of the mass and redshift
dependences can be compressed into a single dependence on the halo 
density bias. Based on this finding, we provide an approximate two-parameter fitting formula. 
\end{abstract}
\maketitle

\section{introduction}
A crucial (but often ignored) assumption in cosmology based on large scale peculiar
velocity is that the galaxy/halo velocity bias $b_v$ equals
unity ($b_v=1$). The usual argument is based on the weak equivalence
principle. 
  On scales larger than about $10$Mpc, gravity is mainly dictated by the
  large scale distribution of dark matter in the Universe, instead of
  individual bound structures such as halos and  galaxies. Therefore the motions
  of halos and  galaxies should faithfully follow this large scale
  structure, leading to $b_v=1$ at $\ga 10$Mpc.

Since galaxies/halos are not the dominant sources of gravity over
$\ga 10$Mpc scale,  they can be treated as test particles. Therefore
their motions should faithfully follow dark matter and therefore
$b_v=1$ at $\ga 10$Mpc scale. 

However, this argument overlooks the fact that halos/galaxies
only reside at (local) density peaks. Along with the fact that the
density gradient is tightly correlated with the velocity field, the
seminal BBKS \citep{BBKS} paper predicted $\sigma_{v,{\rm
    halo}}^2<\sigma_{v,{\rm matter}}^2$. This result was derived using
one-point Gaussian statistics. \citet{Desjacques10} (hereafter DS10)
extended the peak model to 2-point statistics. They derived an elegant
expression, $b_v(k)=1-R_v^2k^2<1$.  The prefactor
$R_v$ depends on the halo mass $M$, but not redshift.  DS10
predicted significant and redshift independent deviation of $b_v$ from
unity, even at scales $\gg 10$ Mpc.   For example,
$b_v\simeq 0.93(0.73)$ at $k=0.1(0.2)\hmpc$ for  
$10^{13}M_{\odot}/h$ proto-halos (e.g. \citet{Elia12}). Later theoretical and numerical works
\citep{Elia12,2012PhRvD..85h3509C,Baldauf14,2015PhRvD..92l3525C}
investigated and verified the DS10 prediction. Nonetheless, {\it these
works  are all for proto-halos defined in the linear and
Gaussian initial conditions, instead of real halos
which host galaxies in observations}. Due to stochastic relation
between proto-halos and halos, and complexities in halo velocity
evolution (e.g. \citet{Colberg00}), ambiguities exist to extrapolate
these works to peculiar velocities of real halos/galaxies. 

Therefore, despite of decades of effort, the velocity bias remains an
unresolved issue.  Even worse, it will become a limiting factor of
peculiar velocity cosmology in the near future
(e.g. \citet{2017MNRAS.464.2517H}). In particular, stage IV dark energy surveys such
as DESI, PFS, Euclid, SKA and WFIRST
(e.g. \citet{BigBOSS,WFIRST15,2015aska.confE..17A,DESI16,Euclid16}) aim to  
measure the structure  growth rate $f(z)\sigma_8(z)$ to $1\%$ or higher
accuracy, through redshift space distortion (RSD). However, what RSD actually measures
is the galaxy peculiar velocity and therefore the combination $b_v\times
f\sigma_8$. Systematic bias in the understanding of $b_v$ then induces a
systematic error in $f\sigma_8$,
\ba
\left.\frac{\delta(f\sigma_8)}{f\sigma_8}\right|_{k,z}\simeq -\left.\frac{\delta b_v}{b_v}\right|_{k,z}\ .
\ea
$b_v$ is in principle dependent of scale $k$.  So the induced
systematic error depends on the small scale cut $k_{\rm max}$ (namely
we only use the velocity information at $k\leq k_{\rm max}$).  Various
cuts have been adopted in peculiar velocity cosmology.  \citet{DESI16} adopts $k_{\rm
  max}=0.2 \hmpc$ in DESI forecast. The eBOSS collaboration adopts
$0.3\hmpc$ in the quasar power spectrum analysis, and  $r\sim 20\mpch$ in
the quasar correlation function analysis
\citep{2018arXiv180102689G,2018arXiv180102891R,2018arXiv180103043Z,2018arXiv180102656H,2018arXiv180103062Z}. Ambitious stage V dark energy
projects \citep{Cosmicvision} can measure RSD to smaller scales, and
$k_{\rm max}=0.5\hmpc$ will further significantly improve cosmological constraints. 
To match the survey capability, we  have to theoretically understand
$b_v$ at $0.1\%$-$1\%$ level accuracy over the range  
$k\leq 0.2$-$0.5\hmpc$.

\begin{table}[t]
	\caption{Five sets of halo mass bins. The mass unit is
          $10^{12} M_\odot/h$. $\langle M \rangle$ is the mean halo
          mass. $N_h$ is the total number of halos in the
          corresponding mass bin. $b_h$ is the halo density bias at
          $k<0.1\hmpc$. }
	\label{tab:sample}
	\begin{center}
		\scriptsize
		\begin{tabular}{@{}r|c|c|c|c}
			\hline\hline
			 Set ID & mass range & $\langle M \rangle$ &
                         $N_h/10^5$ & $b_h(k<0.1\hmpc)$ \\
			\hline
			 $ A1(z=0.0) $ & $>10$ & 37.7 & 7.1  & 1.36 \\
			 $    z=0.5\ $ & $>10$& 29.8 & 5.5  & 1.89 \\
			 $    z=1.0\ $ & $>10$& 23.7 & 3.5  & 2.68 \\
			 $    z=2.0\ $ & $>10$ & 17.7 & 0.88 & 4.98 \\
			\hline
			\hline
			 $ A2(z=0.0) $ & 1-10 & 2.7 & 54.4 & 0.81 \\
			 $    z=0.5\ $ & 1-10 & 2.6 & 52.5 & 1.04 \\
			 $    z=1.0\ $ & 1-10 & 2.5 & 46.9 & 1.48 \\
			 $    z=2.0\ $ & 1-10 & 2.3 & 28.6 & 2.64 \\
			\hline 
			\hline 
			 $ A3(z=0.0) $ & 0.5-1 & 0.70 & 52.9 & 0.70 \\
			 $    z=0.5\ $ & 0.5-1 & 0.70 & 54.0 & 0.86 \\
			 $    z=1.0\ $ & 0.5-1 & 0.69 & 52.4 & 1.15 \\
			 $    z=2.0\ $ & 0.5-1 & 0.69 & 39.7 & 1.97 \\
			 $    z=3.0\ $ & 0.5-1 & 0.68 & 21.8 & 3.15 \\
			\hline 
			\hline 
			 $ B1(z=0.0) $ & 0.5-1   & 0.70  & 52.9 & 0.70 \\
			 $ B2(z=0.0) $ & 1-7     & 2.4  & 51.5 & 0.80 \\
			 $ B3(z=0.0) $ & 7-10    & 8.2  & 3.3  & 1.02 \\
			 $ B4(z=0.0) $ & $>10$   & 37.7 & 7.1  & 1.36 \\
			\hline 
			\hline 
			 $ C1(z=0.0) $ & 7-10      & 8.2  & 3.3  & 1.02 \\
			 $ C2(z=0.5) $ & 1.2-4.0   & 2.1  & 34.1 & 1.02 \\
			 $ C3(z=1.0) $ & 0.31-0.35 & 0.33 & 20.3 & 1.01 \\
			\hline 
		\end{tabular}
	\end{center}
\end{table}
Since halos are highly nonlinear objects, we
have to resort to high precision cosmological  simulations to
accurately measure their velocity statistics. State of art simulations
are already able to reliably simulate the formation and evolution of
halos hosting target galaxies in cosmological surveys, and generate
accurate phase-space distribution of halos.

Nevertheless, translating the accurately simulated halo
phase-space distribution into the volume weighted velocity
statistics\footnote{Unlike the density weighted velocity statistics,
  the volume weighted velocity does not depend on the galaxy
  density bias.  For theoretical modeling of RSD, the volume 
weighted velocity statistics is 
  preferred in some approaches
  (e.g. \citet{Kaiser87,Scoccimarro04,Taruya10,Zhang13}), while the
  density weighted statistics is preferred in the
  distribution function approach 
  \citep{Seljak11,2012JCAP...11..014O} and the streaming model \citep{Peebles80,White15}. } is highly non-trivial, due to a long-standing
problem of ``sampling artifact''.  This problem exists for galaxies, halos
and simulation particles. One way to demonstrate its existence is to randomly
select a fraction $f_{\rm sub}$ of these objects and then measure the velocity
power spectrum of this sub-sample.  The measured
velocity power spectrum should be independent of $f_{\rm sub}$. However,
both theoretical and numerical investigations show significant
dependence, and therefore the existence of sampling artifact \citep{Zhang15a,Zheng15a}.  This sampling artifact problem
arises from the  fact  that we only know the 
velocities where there are objects (halos, galaxies, simulation
particles). Their distributions are not only inhomogeneous, but also
correlated with their velocity fields. So the
sampling of their velocity fields is biased, leading  to biased
measurement of volume weighted velocity statistics 
(e.g. \citet{DTFE96,Bernardeau97,DTFE00,Pueblas09,Zheng13,Zhang15a,Zheng15a,2015MNRAS.446..793J}). 

\begin{table}[t]
	\caption{The determined velocity bias. We discover
          statistically significant deviation of $b_v$ from unity at
          $k\geq 0.1\hmpc$. $|b_v-1|$ increases with $k$, and  may  reach
          $\mathcal{O}(10\%)$ at $k\sim 0.3\hmpc$. } 
	\label{tab:velbias}
	\begin{center}
		\scriptsize
		\begin{tabular}{@{}r|c|c|c}
			\hline\hline
			 Set ID & $(b_v-1)\times 100$& $(b_v-1)\times 100$& $(b_v-1)\times 100$\\
              & $0.05<k<0.1$ & $0.15<k<0.2$ & $0.25<k<0.3$\\
			\hline
			 $ A1(z=0.0) $ &$ 0.03 \pm 0.13$ &$-0.05 \pm 0.38$ &$-0.14 \pm 0.93$\\
			 $    z=0.5\ $ &$-0.02 \pm 0.16$ &$-0.29 \pm 0.32$ &$-1.01 \pm 0.63$\\
			 $    z=1.0\ $ &$-0.04 \pm 0.27$ &$-0.40 \pm 0.53$ &$-2.57 \pm 0.97$\\
			 $    z=2.0\ $ &$-0.31 \pm 0.43$ &$-1.46 \pm 0.82$ &$-6.90 \pm 1.32$\\
			\hline
			\hline
			 $ A2(z=0.0) $ &$ 0.04 \pm 0.06$ &$ 0.37 \pm 0.12$ &$ 1.21 \pm 0.33$\\
			 $    z=0.5\ $ &$ 0.07 \pm 0.06$ &$ 0.27 \pm 0.14$ &$ 0.81 \pm 0.19$\\
			 $    z=1.0\ $ &$ 0.04 \pm 0.06$ &$ 0.06 \pm 0.08$ &$ 0.26 \pm 0.20$\\
			 $    z=2.0\ $ &$-0.05 \pm 0.08$ &$-0.37 \pm 0.18$ &$-1.25 \pm 0.13$\\
			\hline 
			\hline 
			 $ A3(z=0.0) $ &$-0.01 \pm 0.06$ &$ 0.40 \pm 0.15$ &$ 1.50 \pm 0.19$\\
			 $    z=0.5\ $ &$ 0.05 \pm 0.04$ &$ 0.35 \pm 0.13$ &$ 1.06 \pm 0.21$\\
			 $    z=1.0\ $ &$ 0.04 \pm 0.05$ &$ 0.22 \pm 0.15$ &$ 0.60 \pm 0.17$\\
			 $    z=2.0\ $ &$-0.04 \pm 0.07$ &$-0.08 \pm 0.15$ &$-0.31 \pm 0.15$\\
			 $    z=3.0\ $ &$-0.12 \pm 0.10$ &$-0.60 \pm 0.22$ &$-1.58 \pm 0.26$\\
			\hline 
			\hline 
			 $ B1(z=0.0) $ &$-0.01 \pm 0.06$ &$ 0.40 \pm 0.15$ &$ 1.50 \pm 0.19$\\
			 $ B2(z=0.0) $ &$ 0.04 \pm 0.07$ &$ 0.33 \pm 0.14$ &$ 1.22 \pm 0.24$\\
			 $ B3(z=0.0) $ &$-0.01 \pm 0.28$ &$ 0.43 \pm 0.50$ &$ 0.57 \pm 1.03$\\
			 $ B4(z=0.0) $ &$ 0.03 \pm 0.13$ &$-0.05 \pm 0.38$ &$-0.14 \pm 0.93$\\
			\hline 
			\hline 
			 $ C1(z=0.0) $ &$-0.01 \pm 0.06$ &$0.43 \pm 0.50$ &$ 0.57 \pm 1.03$\\
			 $ C2(z=0.5) $ &$ 0.06 \pm 0.07$ &$0.34 \pm 0.17$ &$ 0.88 \pm 0.38$\\
			 $ C3(z=1.0) $ &$ 0.04 \pm 0.07$ &$0.29 \pm 0.22$ &$ 0.66 \pm 0.35$\\
			\hline 
		\end{tabular}
	\end{center}
\end{table}

For 
the dark matter velocity statistics, this is essentially an issue of mass
resolution. If on the average there are $\gg 10$ simulation particles
in a volume $L^3$, the velocity field above the scale $L$ is well
sampled and the sampling artifact is negligible.  Increasing the mass
resolution increases the number density of simulation particles and
pushes the reliable measurement to smaller scales. Unfortunately, the sampling
artifact in the halo velocity statistics is much worse and can not be
reduced by increasing the mass resolution. First, halos
are much more sparse than simulation particles,  causing much severer
sampling artifact. Second, increasing the mass resolution can not
alleviate the sampling artifact  problem, since the halo number density is
fixed at given redshift and mass. This sampling artifact prohibits
accurate determination of halo velocity bias in simulations, with existing
methods. 

We have tried various approaches to measure the volume weighted
velocity statistics in simulations.  We have designed new
velocity assignment methods, including the NP method \citep{Zheng13} and the
Kriging method \citep{Yu15,Yu17}. We have built theoretical
model of the sampling artifact \citep{Zhang15a},
verified it in simulated dark matter velocity field \citep{Zheng15a} and
then applied it to correct the sampling artifact in the halo velocity field
\citep{Zheng15b}.  Despite these efforts, we have not yet succeeded in measuring
$b_v$ with $1\%$ accuracy at $k=0.1\hmpc$. The $b_v$ measurement at
larger $k$ is even more challenging.

The current paper presents an exact approach to determine $b_v$ from simulations. It
circumvents the problem of sampling artifact by a hybrid method. For the first time, we determine $b_v(k,z)$ to $0.1\%$-$1\%$ 
at $k\leq 0.4\hmpc$ and $0<z<2$, for various halo mass bins. In \S
\ref{sec:method}, we present our method of determining $b_v$. We leave
further technical details in the appendix.  In \S
\ref{sec:simulation} we describe the simulation used for the velocity
bias measurement. In \S \ref{sec:result} we present the determined
$b_v(k)$ for various halo mass and redshift. In \S \ref{sec:survey} we
discuss its impact on cosmological surveys. We discuss and conclude
in \S \ref{sec:conclusion}.

\section{The method}
\label{sec:method}
The method is hybrid in the sense that it is composed of two steps,
direct measurements and theoretical interpretation. 
\bi
\item Direct measurements are for three quantities, all with negligible sampling
artifact. (1) One is the  halo momentum ${\bf p}_h\equiv (1+\delta_h){\bf
  v}_h$. (2) One is the halo number overdensity $\delta_h({\bf x})$, for which
the sampling artifact is irrelevant.  It is the density weighted velocity, free of sampling
artifact. (3) One is the matter velocity ${\bf v}_m$. In principle it
contains sampling artifact. Fortunately, it has been suppressed to a negligible level
in our simulation. Our simulation has on the average $216$ simulation
particles per $(2.4\mpch)^3$ volume.  For such dense sampling, the resulting sampling
  artifact in the dark matter velocity power spectrum is
  $0.02\%\times (k/(0.1\hmpc))^2$
  (estimated by Eq.16\&24 in \citet{Zhang15a}). The induced
systematic error in the determined $b_v$ is half of that, $0.01\%\times
(k/(0.1\hmpc))^2$. Therefore the dark matter
velocity field at scales of interest ($k\la 0.4\hmpc$)  is well
sampled and is essentially free of sampling artifact. 
\item These direct measurements are then exactly linked to  the volume weighted
  statistics in theory,  where the only unknown parameter is
  $b_v(k)$. We then solve for $b_v$. The determination of $b_v$ is
  then free of sampling artifact. 
\ei

\bfi{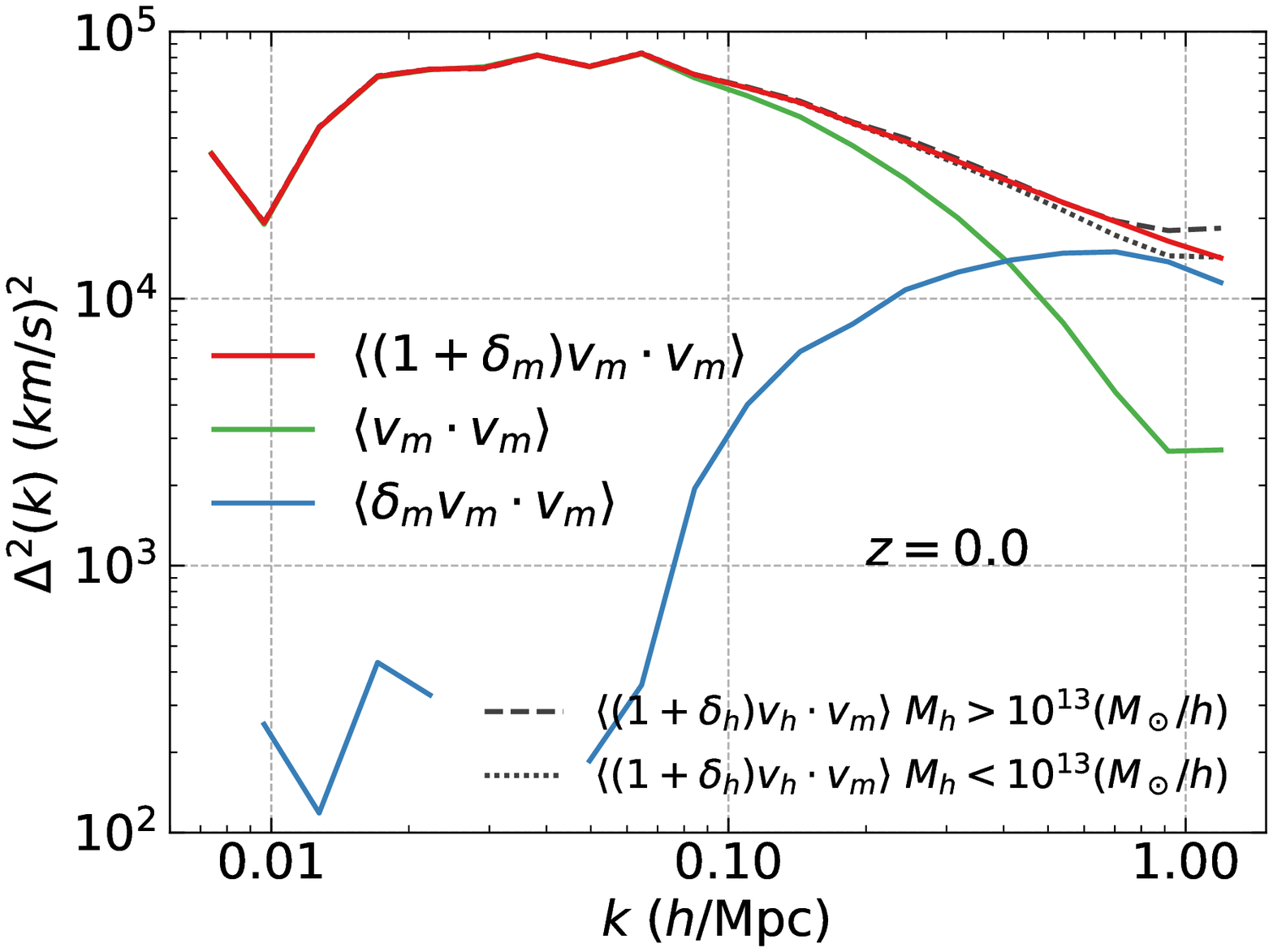}
\caption{The $z=0$ power spectrum variance $\Delta^2_\alpha\equiv
  k^3P_{\alpha}(k)/2\pi^2$ in unit of $({\rm km}/s)^2$, where $\alpha=(1+\delta_m)v_mv_m$,
  $(1+\delta_h)v_hv_m$, $v_mv_m$, $\delta_mv_mv_m$. All these
  measurements are essentially free of the sampling artifact in 
  the velocity field. At $k<0.3\hmpc$, 
  $P_{(1+\delta_m)v_mv_m}$ is dominated by $P_{v_mv_m}$. This property
  makes the measurement of $b_v$ easier.  \label{fig:measurement}}
\efi

{\bf Step 1}. We first measure the correlation function 
\ba
\xi_{(1+\delta_h)v_hv_m}(r)\equiv \langle (1+\delta_h({\bf x}_1)){\bf v}_h({\bf x}_1)\cdot {\bf v}_m({\bf x}_2)\rangle
\ea
and its Fourier counterpart $P_{(1+\delta_h)v_hv_m}(k)$. These
measurements are then linked to the following 
correlation functions
\ba
\label{eqn:correlation}
\xi_{(1+\delta_h)v_hv_m}(r)=\langle{\bf v}_h\cdot {\bf v}_m\rangle+\langle
\delta_h{\bf v}_h\cdot {\bf v}_m\rangle
\ea
and their power spectra (Fourier components) 
\ba
\label{eqn:P1}
P_{(1+\delta_h)v_hv_m}(k)&=&P_{v_hv_m}(k)+B_{\delta_hv_hv_m}(k)\ .
\ea
Here $\langle \cdots \rangle$ denote the volume/ensemble averaging. 

{\bf Step 2}. We then solve Eq. \ref{eqn:P1} for the velocity bias
$b_v(k)$, defined through 
\ba
\label{eqn:bv}
b_v(k)\equiv \frac{P_{v_hv_m}(k)}{P_{v_mv_m}(k)}\ .
\ea
A key point to pay attention is that $b_v(k)$ is the only unknown
quantity in Eq. \ref{eqn:P1}. The proof is presented in the appendix. 
Then comparing the left and right hand sides drawn from the same simulation, we determine
$b_v$. For this reason, the determined $b_v$ is essentially free of cosmic variance. In the appendix, we present the maximum likelihood
approach to solve Eq. \ref{eqn:P1} for $b_v(k)$.  

\bfi{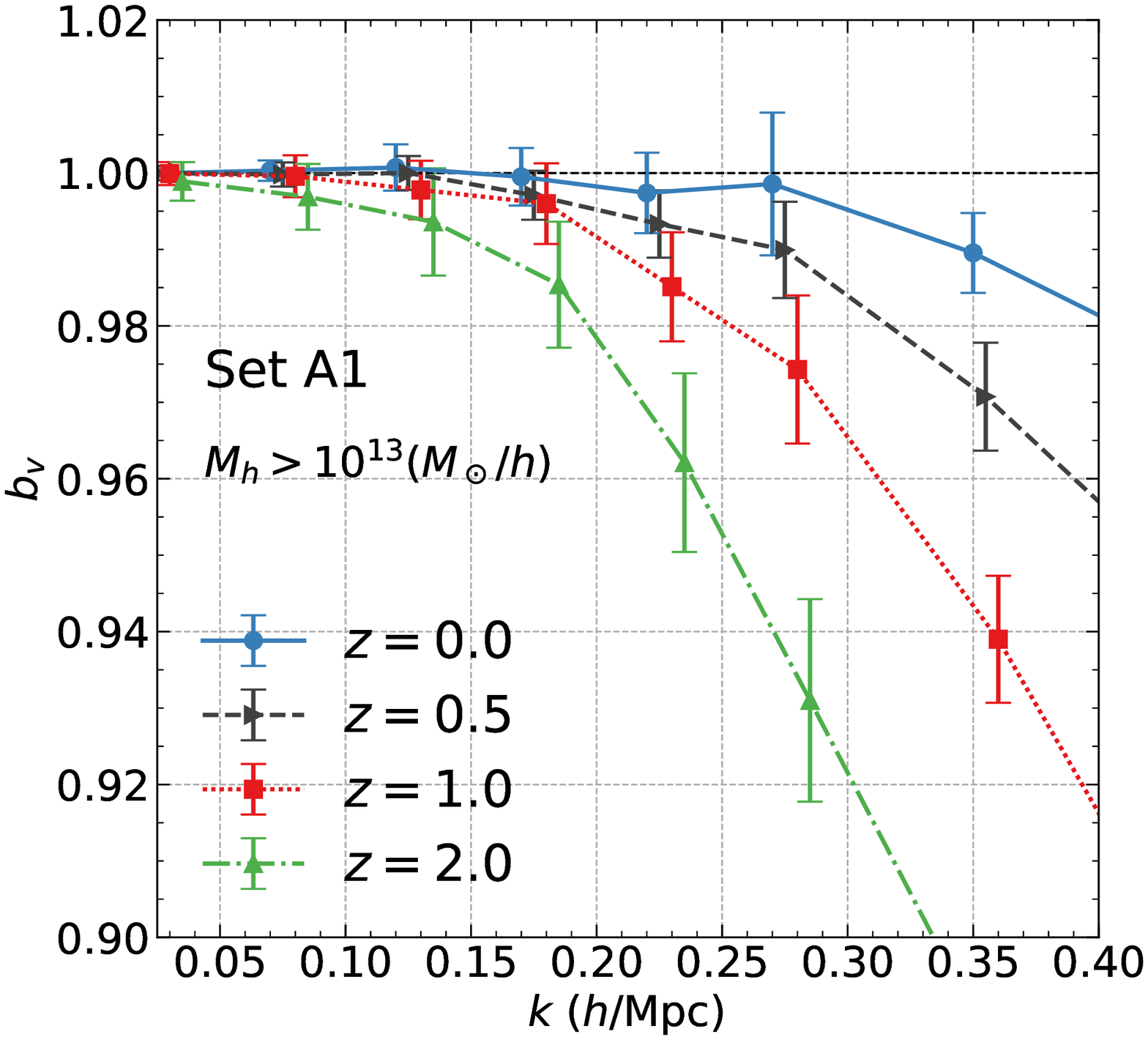}
\caption{The velocity bias of halo set A1 ($M>10^{13} M_{\odot}/h$) at $z=0,
  0.5, 1.0, 2.0$. The velocity bias decreases with increasing
  redshift. For these halos, $b_v<1$ at all redshifts. Notice that
  for clarity we shift the $z>0$ data points horizontally. The result
  invalidates the usual assumption of $b_v=1$  in peculiar velocity
  cosmology. $|b_v-1|$ is much weaker than the peak model
  prediction. It also shows significant redshift evolution, in
  contrast to the peak model prediction.   \label{fig:A1}}
\efi
\section{The simulation}
\label{sec:simulation}
The simulation we analyze (J6620) adopts the standard  $\Lambda$CDM
cosmology, with  $\Omega_m=0.268$, $\Omega_\Lambda=0.732$,
$\Omega_b=0.044$,  $\sigma_8=0.83$,  $n_s=0.96$ and  $h=0.71$.  It has
box size $L_{\rm box}=1200{\rm Mpc}/h$, particle number $N_P=3072^3$
and the mass resolution $4.4\times 10^{9} M_\odot$. J6620 is run with
a particle-particle-particle-mesh (${\rm P^3M}$) code, detailed in
\cite{Jing07}. The halo catalogue is constructed by the Friends-of-Friends
(FOF) method. The linking length is $b=0.2$. In the catalogue gravitationally unbound
``halos'' have been excluded. The halo center is defined as the mass
weighted center and the halo velocity is defined as the velocity averaged over
all member particles. We adopt
various mass and redshift bins to calculate the mass and redshift dependence of
velocity bias. Table \ref{tab:sample} shows details of these
bins.

\bfi{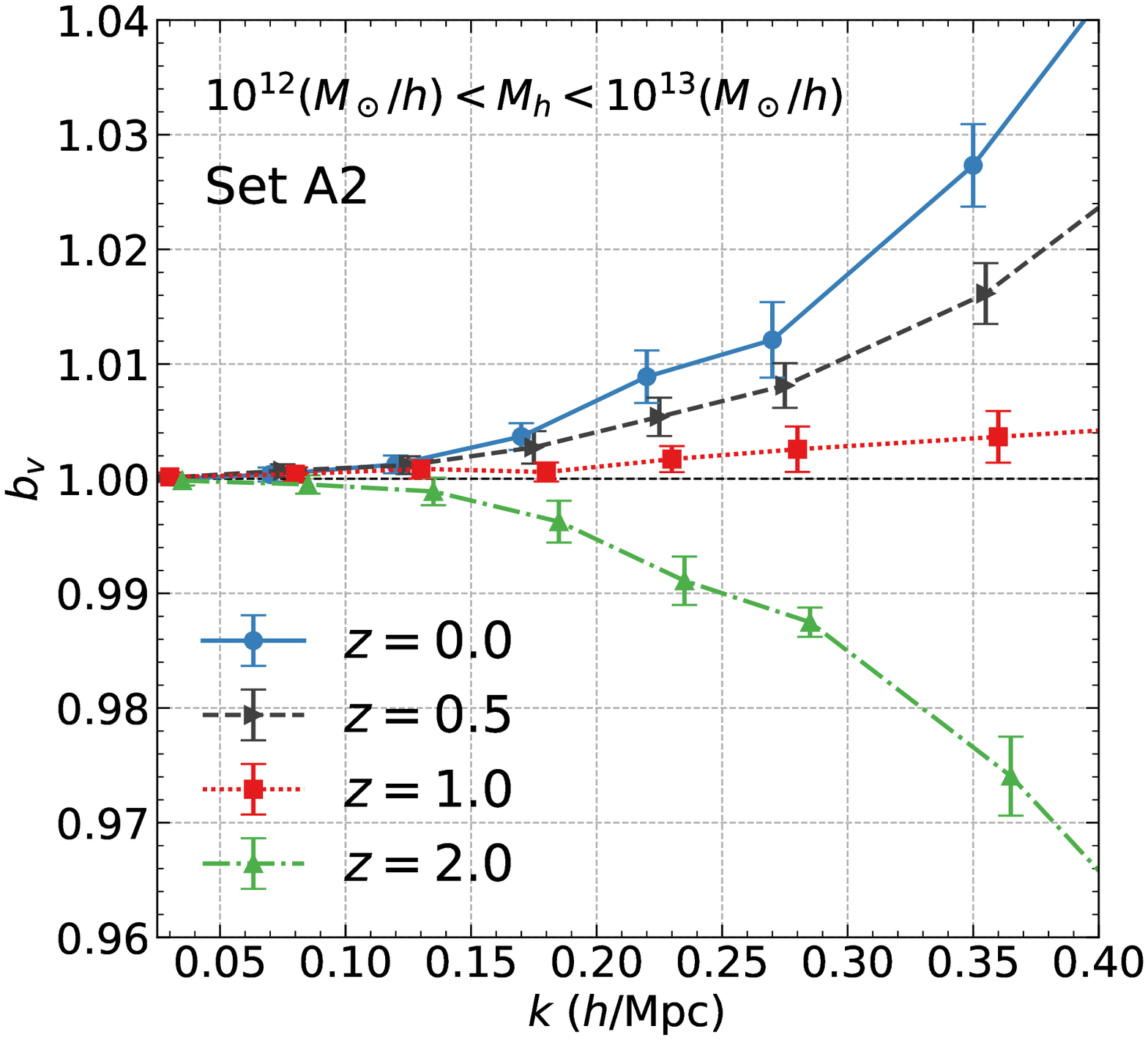}
\caption{Similar to Fi.g \ref{fig:A1}, but for halo set A2
  ($10^{12}<M/(M_{\odot}/h)<10^{13}$). Notice that $b_v-1$ changes
  from positive sign to negative sign from $z=0$ to $z=2$. \label{fig:A2}}
\efi

The number density and momentum fields of both halos and dark matter are  measured using
the NGP method. Namely $(1+\delta_h){\bf 
  v}_h=\sum_i {\bf v}^i_h/\bar{n}$. The sum ($\sum_i$) is over all particles in
the given cell. $\bar{n}$ is the mean number
of halos in each cell. We adopt $512^3$ grid points. The grid cell
size is $L_{\rm grid}=2.4\mpch$. Each cell has 216 simulation particles on the
average. Therefore we have excellent sampling of the dark matter velocity
field above such scale. This allows us to robustly measure the dark
matter velocity field through ${\bf v}_m=\sum_i {\bf v}_m^i/\sum_i$,
with negligible sampling artifact.   The aliasing effect is also
negligible, since  we are interested in the scales of $k\la 0.4\hmpc$,
much smaller than the Neyquist wavenumber 
$k_N=\pi/L_{\rm grid}=1.31\hmpc$ \citep{Jing05}.

Fig. \ref{fig:measurement} shows the directly measured
$P_{(1+\delta_h)v_hv_m}$ for $M>10^{13}M_\odot/h$ and
$10^{12}M_\odot/h<M<10^{13}M_\odot/h$ halos at $z=0$. For comparison, we also show
$P_{(1+\delta_m)v_mv_m}$.  The three are almost identical to each other until
$k\ga 0.2\hmpc$. These terms have two contributions. The contribution
from $\langle {\bf v}\cdot{\bf v}\rangle$ dominates at $k\la
0.3\hmpc$. The contribution from $\langle \delta {\bf v}\cdot{\bf
  v}\rangle$ becomes significant at $k\ga 0.2\hmpc$ and becomes
dominant at $k\ga 0.3\hmpc$. These results already imply $b_v\simeq1$ at
$k\la 0.1\hmpc$. The exact determination of $b_v(k)$ requires to solve
Eq. \ref{eqn:P1} using the method described 
in the appendix.

\section{The velocity bias}
\label{sec:result}
We obtain the best-fit value and the associate error of $b_v(k)$ over
the $k$ ranges of $(0,0.05)$, $(0.05,0.1)$, $(0.1,0.15)$, $(0.15,0.20)$,
$(0.20,0.25)$, $(0.25,0.3)$, $(0.3,0.4)$, $(0.4,\infty)$.  As explained
early, the determined
$b_v$ is essentially free of cosmic variance, since it is obtained by
comparing the halo and dark matter fields in the same simulation
box. The only source of noise is shot noise in the halo
distribution. The large number of halos ($\sim 10^{5-7}$) then enables
us to determine $b_v(k\la 0.4\hmpc)$ with $0.1\%$-$1\%$
statistical error. Such accuracy also enables us to detect any significant deviation of $b_v$
from unity.  

\bfi{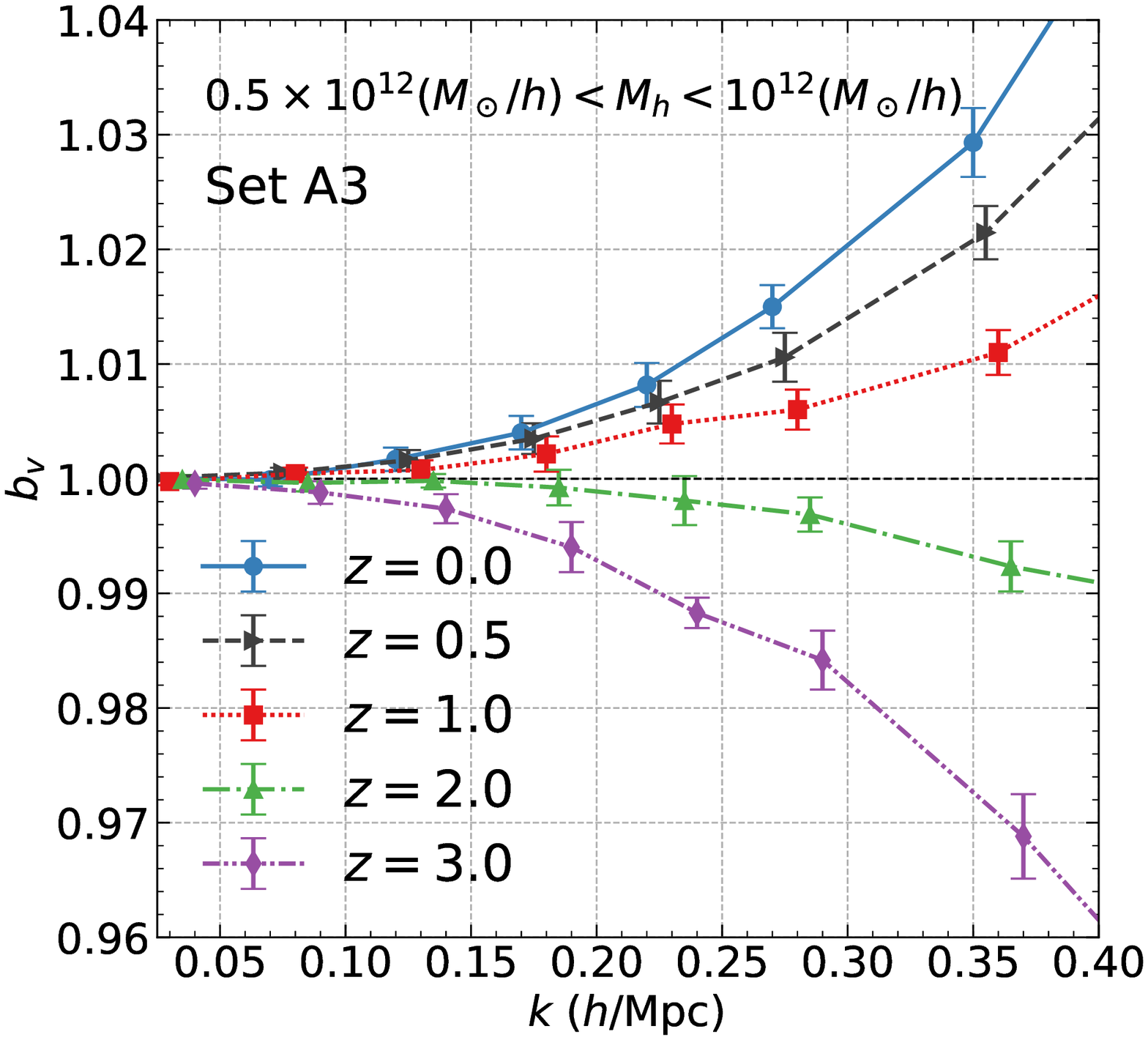}
\caption{Similar to Fi.g \ref{fig:A1}, but for set A3 ($5\times
  10^{11}<M/(M_{\odot}/h)<10^{12}$). Again, $b_v-1$ changes sign with redshift. \label{fig:A3}}
\efi

Fig. \ref{fig:A1}, \ref{fig:A2} \& \ref{fig:A3} show the redshift
dependence of velocity bias for three mass bins (halo set A1, A2, A3 respectively). Fig. \ref{fig:z0} shows the mass dependence of velocity
bias at $z=0$ (set B). Furthermore, table \ref{tab:velbias}
lists the velocity bias at selected ranges of $k$. 

We detect statistically significant deviation of $b_v$ from
unity at $k\geq 0.1\hmpc$. This invalidates the assumption of $b_v=1$
commonly adopted in peculiar velocity cosmology. It will significantly
impact RSD cosmology of stage IV dark energy projects. The
deviation $b_v-1$ shows rich behavior in $k$, halo mass $M$ and
redshift $z$. Nonetheless, it shows significant difference to the peak model
prediction and poses new question to the halo peculiar velocity theory.  Major findings are as follows. 

\subsection{The $k$ dependence}
$b_v(k)-1$ can have either negative or positive sign. This challenges
the peak mode, which predicts a negative sign. The  sign of $b_v-1$ does
not vary with $k$. Furthermore, $|b_v(k)-1|$ increases with $k$, and
roughly speaking $b_v(k)-1\propto k^2$. (1) At $k\leq 0.1\hmpc$, the 
deviation is very weak. Over all the halo mass and
  redshift investigated, the deviation is $0.3\%$ or less and it is
  statistically insignificant. $|b_v-1|$ is orders of magnitude weaker than the peak
  model prediction on proto-halos. It means that we can not simply
  extrapolate the predictions on proto-halo velocity to real halo
  velocity.  (2) At $0.1\leq k\leq 0.2\hmpc$, $b_v$ may show statistically
  significant deviation from unity.  Depending on the type of halos,
  the deviation may reach $1\%$. As discussed later, despite its
  weakness, it will become a significant source of systematic error
  for DESI RSD cosmology.  (3) At $0.2\leq k\leq 0.4\hmpc$,  some halo
  samples show $\mathcal{O}(10\%)$ deviation from unity.  One task of
  RSD cosmology is to extract cosmological information deep into the nonlinear regime of $k\leq
  0.5\hmpc$ (e.g., the cosmic vision dark energy report
  \citep{Cosmicvision}). The existence of significant deviation of
  $b_v$ from unity at this regime is a challenge to this task.

\bfi{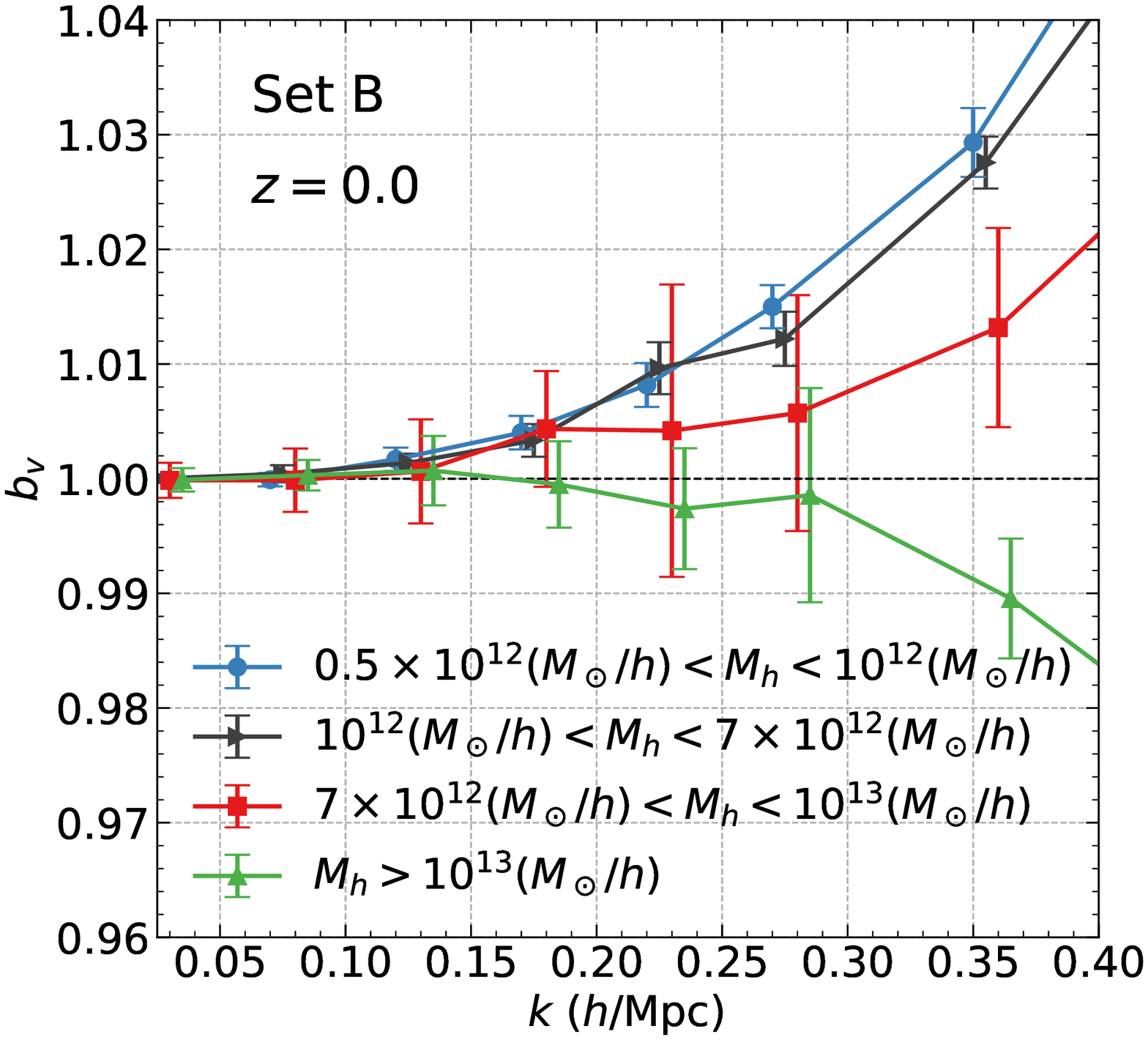}
\caption{The dependence of halo velocity bias on halo mass (halo set B at $z=0$). For clarity,
  we shift the results of the three higher mass bins horizontally.
  The sign of $b_v-1$ not only changes with redshift, but also with
  mass.  \label{fig:z0}}
\efi

\subsection{The mass and redshift dependence}
$b_v$ increases with  increasing
  redshift (Fig. \ref{fig:A1}, \ref{fig:A2} \& \ref{fig:A3}) and
  decreasing mass (Fig. \ref{fig:z0}).  As a consequence, the sign of  $b_v-1$ depends on halo mass
  and redshift. For example, $b_v-1$ is always negative for halos of
  $M>10^{13}M_{\odot}/h$ at all redshifts (Fig. \ref{fig:A1}). But it
  changes from positive at $z=0$ to negative at $z=2$ for 
 less massive halos (Fig. \ref{fig:A2} \& \ref{fig:A3}). Another
 consequence is that $b_v-1$ has strong dependence on the halo mass and
 redshift. For example,  for $M>10^{13}M_{\odot}/h$  halos at $0.25<k<0.3 \hmpc$ and
$z=1(2)$, $b_v-1=-0.026(-0.069)$. But for halos of
$10^{12}<M<10^{13}M_{\odot}/h$, $b_v-1\sim 0.3\%$ at
$z=1$. Fig. \ref{fig:z0} compares $b_v$ of various mass at $z=0$. Now
the biggest deviation from unity happens for the least massive
halos. 

To translate the above results into impact on peculiar velocity  cosmology, we need
specifications of galaxy surveys,  since different surveys probe different
galaxies in different halos and different redshifts. Here we just
present a qualitative description on halos that may be probed by
various surveys.  Later will quantify the impact of velocity bias to
some of these surveys in \S \ref{sec:survey}.  (1) 
$10^{13}M_\odot$ halos at $z<1$ may be probed by LRGs in DESI
(e.g. \citet{Guo15,DESI16}). Galaxies in the TAIPAN redshift and
peculiar velocity survey are also expected to reside in these halos, but at $z\sim 0$ 
\citep{2017MNRAS.464.2517H}. (2) For smaller halos ($\la
10^{13}M_\odot$), 21cm surveys may  be capable of detecting them. SKA
are capable of detecting billions of 21cm emitting galaxies residing
in these  
halos, given its sensitivity in HI mass
\citep{2005MNRAS.360...27A,YangXJ11} and the observationally
constrained HI mass-halo mass relation (e.g. \citet{Guo17}). SKA is
also able to indirectly detect them through the intensity
mapping. (3)
For halos of $M<10^{12}M_\odot$, a large fraction of ELGs in surveys such as DESI
and PFS may reside in these halos
\citep{2016MNRAS.461.3421F,2017arXiv170807628G}. DESI can probe them at $0.6\leq z\leq
1.6$ while PFS can probe them to higher redshift ($z\leq 2.4$). 21cm
intensity mapping surveys such as CHIME \citep{CHIME14} and  Tianlai
\citep{Tianlai15} are also sensitive to these halos, although they may not be able to detect
individual galaxies.


\bfi{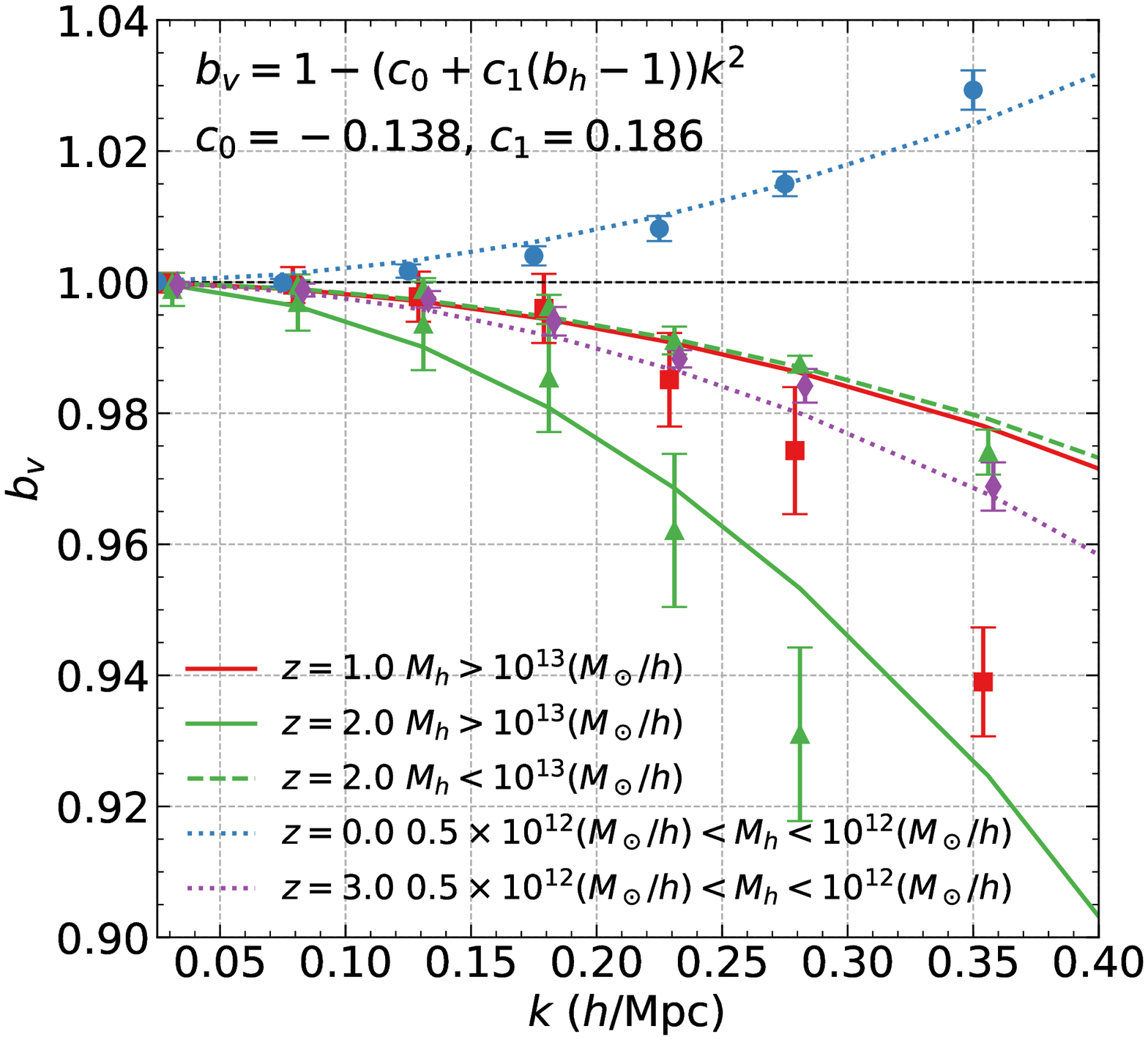}
\caption{A fitting formula of $b_v$. For clarity, we do not show the
  results of all halo sets.  The residual error mainly arises from the
  dependence of the velocity bias $b_v$ beyond the density bias $b_h$.
  It is an issue for further investigation.  \label{fig:fit}} 
\efi

\subsection{The dependence on the halo density bias}
An interesting observation is that the mass and redshift dependence of
$b_v$ may be absorbed into a single dependence on the halo density bias
$b_h$.  This can be seen by first checking $b_h$ in the table
\ref{tab:sample} and then comparing $b_v$ of halos with similar
$b_h$. More explicitly, halos of set C (Table \ref{tab:sample} ) at
different redshifts are chosen to have identical density
bias ($\simeq 1$). Table. \ref{tab:velbias} shows that they have roughly the same
velocity bias.  This motivates us to propose the following fitting
formula,
\ba
\label{eqn:fit}
b_v(k|M,z)\simeq 1-\left[c_0+c_1(b_h(M,z)-1)\right]\tilde{k}^2\ . 
\ea
Here $\tilde{k}\equiv k/(\mpch)$. This is basically the Taylor
expansion of $b_v({\bf k})$ around $(0,0,0)$. The isotropy of  the velocity bias ($b_v({\bf
  k})=b_v(k))$) requires that terms of odd power in $k$ such as  $k_m$ and $k_mk_nk_l$
($m,n,l=1,2,3$) vanish in the Taylor expansion. Therefore the leading order term is
$k^2$.  We find that $c_0=-0.138\pm 0.01$ and $c_1=0.186\pm 0.007$
(Fig. \ref{fig:fit}). Small error in $c_1$ shows that the dependence on $b_h$ is
statistically significant. We caution that this fitting formular is only
approximate, since it ignores dependence beyond $b_h$ and ignores
higher order $k$ dependence (e.g. $k^4$). Nevertheless, it is sufficient to
describe the over dependence of the velocity bias on halo property and
scale (Fig. \ref{fig:fit}).  Another caveat in this fitting
  formula is the implicit assumption that all the
cosmological dependences are encoded in the cosmological dependence of
$b_h$ and therefore $c_{0,1}$ do not depend on cosmology. This is
motivated by the primary dependence of $b_v$ on $b_h$. If valid, we
are then able to use this fitting formular for other cosmologies such
as the Planck 2015 cosmology \citep{2016A&A...594A..13P}, whose $\Omega_m$ is $13\%$
larger. Future works will use simulations of different cosmologies to
investigate this assumption.

\section{Implications for peculiar velocity 
  surveys}
\label{sec:survey}
We discuss two implications of velocity bias on cosmology. The first
is that it may bias the structure growth rate measurement in
spectroscopic redshift surveys. The second is that it may open a window
of testing the equivalence principle at cosmological scales. 

\bfi{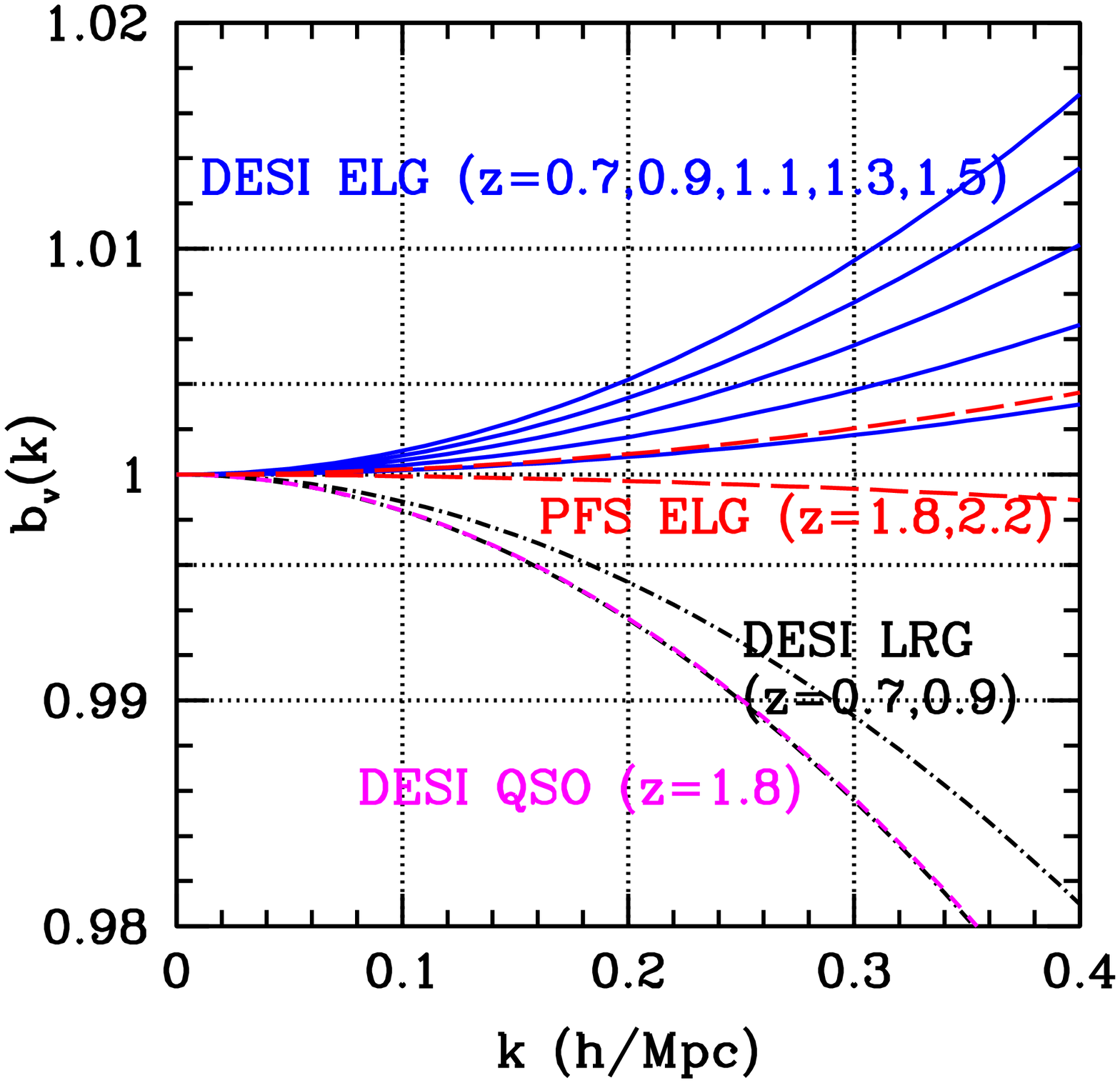}
\caption{Implications for RSD surveys. Notice that the line of DESI
  LRGs at $z=0.9$ almost overlaps with that of DESI QSO at
  $z=1.8$. (1) The dot lines of $1\pm 0.4\%$
  denote the expect overall statistical error in $f\sigma_8$
  constrained by DESI. For DESI,  velocity bias is a source
  of significant systematic errors. (2) The velocity bias will be more significant for
  Euclid, SKA HI survey and the proposed stage V billion 
  object spectra survey, due to their better constraining power in $f\sigma_8$. (3) It is less significant for PFS due to its
  larger statistical error ($\sim 1\%$).  Since $b_v$
  of PFS ELGs at $z<1.6$ is similar to that of DESI ELGs, 
  we only show the PFS results at $z=1.8$ and $2.2$. \label{fig:survey}}
\efi

\subsection{Impact on structure growth rate constraint}
A major task of cosmological surveys is to constrain the structure
growth rate through peculiar velocity. The velocity bias, if ignored
or modelled inappropriately, will become a source of systematic error.
Whether it is of statistical significance depends on
surveys and galaxy types. The low redshift TAIPAN survey aims to
measure the peculiar velocities of $\sim 10^4$ galaxies. It will constrain 
$f\sigma_8$ at $z\sim 0$ with $\sim 10\%$ accuracy, using information at $k<0.2\hmpc$
\citep{2017MNRAS.464.2517H}. The target galaxies (with $M\sim
10^{13}M_\odot/h$) have $|b_v-1|\ll 1\%$ at $k<0.2\hmpc$
(Fig. \ref{fig:A1} and Table \ref{tab:velbias}). Therefore
the usual assumption of $b_v=1$ results into negligible systematic
error, and can be adopted safely.  

On the other hand, the spectroscopic redshift survey DESI can
measure $f\sigma_8$ to $1\%$ level over a number of redshift
bins, and resulting into an overall $0.4\%$ statistical error
\citep{DESI16}. Fig. \ref{fig:survey} plots  the predicted $b_v$ of
various galaxies in DESI.  We estimate $b_v$ using the fitting formula
of Eq. \ref{eqn:fit}. The density biases of various galaxy types are
adopted as $b_{\rm 
  LRG}=1.7/D(z)$,  $b_{\rm ELG}=0.84/D(z)$ and $b_{\rm
  QSO}=1.2/D(z)$ \citep{DESI16}. Here $D(z)$ is
the linear density growth rate and it is normalized as unity at
$z=0$.   For DESI, we are no longer able to approximate
$b_v=1$, otherwise systematic error at $1\sigma$ level can be
induced. 

For PFS ELGs, the predicted $b_v$ at $z<1.6$ is similar to that of
DESI ELGs. So for clarity we neglect them 
  in Fig. \ref{fig:survey}. We only show the results at $z=1.8(2.2)$,
  where we adopt $b_{\rm ELG}=1.62(1.78)$ (the PFS SSP
  proposal). If only using information at $k\leq 0.2\hmpc$,the
  systematic error in $f\sigma_8$ caused by ignoring $b_v\neq 1$ is by less than $0.4\%$. Due to a factor of $10$ smaller sky coverage than DESI, the overall
  statistical error of $f\sigma_8$ constrained by PFS RSD to
  $k=0.2\hmpc$ is expect to be $\sim 1\%$ . Therefore the impact of
  velocity bias on PFS RSD is subdominant and we can simply
  approximate $b_v=1$. However, due to higher 
  number density of PFS galaxies,  it can measure RSD to smaller scales and has the potential to
  further reduce the statistical error in
  $f\sigma_8$. Fig. \ref{fig:survey} shows that,  if we 
  push to $k_{\rm max}=0.3\hmpc$,  we may no longer adopt $b_v=1$. 

The proposed SKA HI survey has the capability of detecting $\sim 10^9$ 21cm emitting
galaxies to $z\la 2$  over $30000$ deg$^2$
\citep{2015aska.confE..17A}. The statistical error in
$f\sigma_8$ is $\sim 0.3\%$ for each $\Delta z\sim 0.1$ redshift bins
over $0.4<z<1.3$. If assuming $b_v=1$, the  induced systematic error
will overwhelm the statistical error. This will also be true for the
proposed stage V survey of measuring a billion spectra of LSST
galaxies \citep{Cosmicvision}.  The
situation for Euclid may fall between DESI and SKA.

\subsection{A cosmological test of the equivalence principle }
Velocity bias also provides a new test of $\Lambda$CDM cosmology. 
Observationally we are not able to measure the velocity bias (with
respect to dark matter)
directly. However,  we are able to measure the ratio of velocity bias
between two tracers of the large scale structure (LSS). Furthermore, if the
two tracers overlap in space, the measured ratio will be free of
cosmic variance  \citep{McDonald09}.  Our result predicts that
in $\Lambda$CDM, the velocity ratio of two tracers is
\ba
\frac{b_{v,1}(k)}{b_{v,2}(k)}-1&\simeq& b_{v,1}(k)-b_{v,2}(k)\no \\
&\simeq&  -0.19\%
\left(\frac{k}{0.1\hmpc}\right)^2\times (b_{h,1}-b_{h,2})\ .
\ea
The first approximation holds since $b_v=1$ at leading order. 
This weak deviation from unity is a genuine consequence
of the equivalence principle (EP). 
Therefore  if $1\%$ or large deviation at $k<0.2\hmpc$ is detected, it
may be a smoking gun of EP
violation and therefore modifications of general relativity (e.g. \citet{Hui09}).  We
will further investigate this issue in future works.

\section{Discussions and conclusions}
\label{sec:conclusion}
We invent a novel method to determine the volume weighted halo velocity
bias $b_v$. This method is free of the long-standing sampling artifact
problem, which has hindered accurate determination of velocity bias. We apply it to a $3072^3$ particle
simulation and measure  $b_v$ to $k\sim 0.4\hmpc$ with better than
$1\%$ accuracy. Our findings confront both the $b_v=1$ standard
assumption in peculiar velocity data analysis, and the peak model
prediction.  (1) There exists statistically significant deviation of
$b_v$ from unity at $k>0.1\hmpc$. Depending on halo mass, redshift,
$b_v-1$ may reach $\mathcal{O}(1\%)$ at $k\sim 0.2\hmpc$ and
$\mathcal{O}(10\%)$ at $k\sim 0.4\hmpc$. If ignored, this velocity
bias will become a significant source of systematic error in RSD
cosmology of DESI. Its impacts on SKA HI galaxy survey
and Euclid are even stronger.   (2) However, $|b_v-1|$ is a factor of $\sim 10$ smaller
than the prediction of peak model. Furthermore, its mass and
redshift dependence do not agree with the peak model prediction.
$b_v$ varies with redshift, while the peak model 
predicts the opposite. $b_v$ of less massive halos can be bigger
than unity, while the peak model always
predicts $b_v<1$. The peak model is based on
proto-halo statistics. Therefore we have to consider the mis-match
between proto-halos and real halos, and the displacement of halos from
the corresponding proto-halos, to improve the theoretical
understanding of velocity bias. Another issue to consider is the
displacement of halos from their initial positions (Lagrangian
positions) to the
present positions (Eulerian positions). This affects the velocity
correlation, which is defined in Eulerian space. It is expected to
make $b_v$ larger than the peak model prediction, and make it
increasing towards $z=0$. We also expect that the environment that halos
reside (e.g. filaments or clusters) may play a role in the halo
velocity bias.  For example, the infall velocity within filaments may
be responsible or partly responsible to  the
$b_v>1$ behavior of  less massive halos. 

There are many issues for further investigations. For example, since
the velocity bias depends on the density bias, would it also depend on
the halo formation time? Or more general, besides the density bias,
what could affect the velocity bias? How does it depend on parameters
within the standard cosmology? How does it behave in modified gravity
cosmology? Also, to robustly predict its impact on RSD cosmology, we
need to generate realistic mocks for target surveys and measure the
velocity bias of LRGs, ELGs, 21cm emitting galaxies, etc.

\section*{Acknowledgments}
This work was supported by the National Science Foundation of China
(11621303, 11433001, 11653003, 11320101002, 11533006),  National
Basic Research Program of China (2015CB85701, 2015CB857003). 

\appendix
\section{The algorithm to solve for the scale dependent velocity bias}

Here we describe in details the procedure of solving for the scale
dependent  $b_v(k)$.  The key is the statement in \S \ref{sec:method},
that $b_v(k)$ is the only unknown
quantity in Eq. \ref{eqn:P1}, where all other quantities are provided
by the same simulation. The proof is as follows. In Fourier space,
we can decompose ${\bf v}_h({\bf 
    k})=b_v(k){\bf v}_m({\bf k})+{\bf v}_h^S({\bf k})$. The first term
  is completely correlated with the density velocity field. In contrast, the
  second term ${\bf v}_h^S$ 
  is the stochastic component of halo velocity. It
	is uncorrelated to the 
  density velocity field at two-point level. 
  Namely, $\langle
  {\bf v}^S({\bf k})\cdot {\bf v}^{*}_m({\bf k})\rangle=0$ ($\langle
  {\bf v}^S({\bf x}_1)\cdot {\bf v}_m({\bf x}_2)\rangle=0$). One can
  verify that the above condition leads to Eq. \ref{eqn:bv} as the
  definition of $b_v(k)$. The decomposition above is therefore
  uniquely fixed. Clearly, ${\bf v}_h^S$ does not contribute
  to $P_{v_h v_m}$. Furthermore it 
  does not contribute to 
$B_{\delta_h v_hv_m}(k)$.  The direction of
  $\hat{v}_h^S({\bf x}_1)$ is uncorrelated with $\delta_h({\bf x}_1)$ due to
  the statistical isotropy of the Universe. This holds no matter the
  halo density bias is scale dependent or non-local, otherwise the
  statistical isotropy will be violated. The direction of
  $\hat{v}_h^S({\bf x}_1)$ is also
  uncorrelated with ${\bf v}_m({\bf x}_2)$, by definition. Averaging
  over its direction, we have  $\langle
\delta_h({\bf x}_1){\bf v}^S_h({\bf x}_1)\cdot {\bf v}_m({\bf
  x}_2)\rangle_{\hat{v}_h^S}=0$. Therefore ${\bf v}_h^S$
does not contribute to the right hand side of Eq. \ref{eqn:P1}.  Since
we know $\delta_h({\bf x})$ and ${\bf v}_m({\bf x})$ from the same simulation, $b_v(k)$
  is all we need to fix the right hand
  side of Eq. \ref{eqn:P1}. 

We are then able  to determine $b_v(k)$
uniquely. The only complexity in determining $b_v(k)$ is the
non-local dependence of the right hand side of
Eq. \ref{eqn:P1}  on $b_v(k)$. It is caused by $B_{\delta_h
  v_hv_m}(k)$, in which  $b_v(k^{'}\neq k)$ also contributes. Here we
present the maximum likelihood solution to $b_v(k)$.

\begin{figure}[ht]
\includegraphics[width=0.36\textwidth]{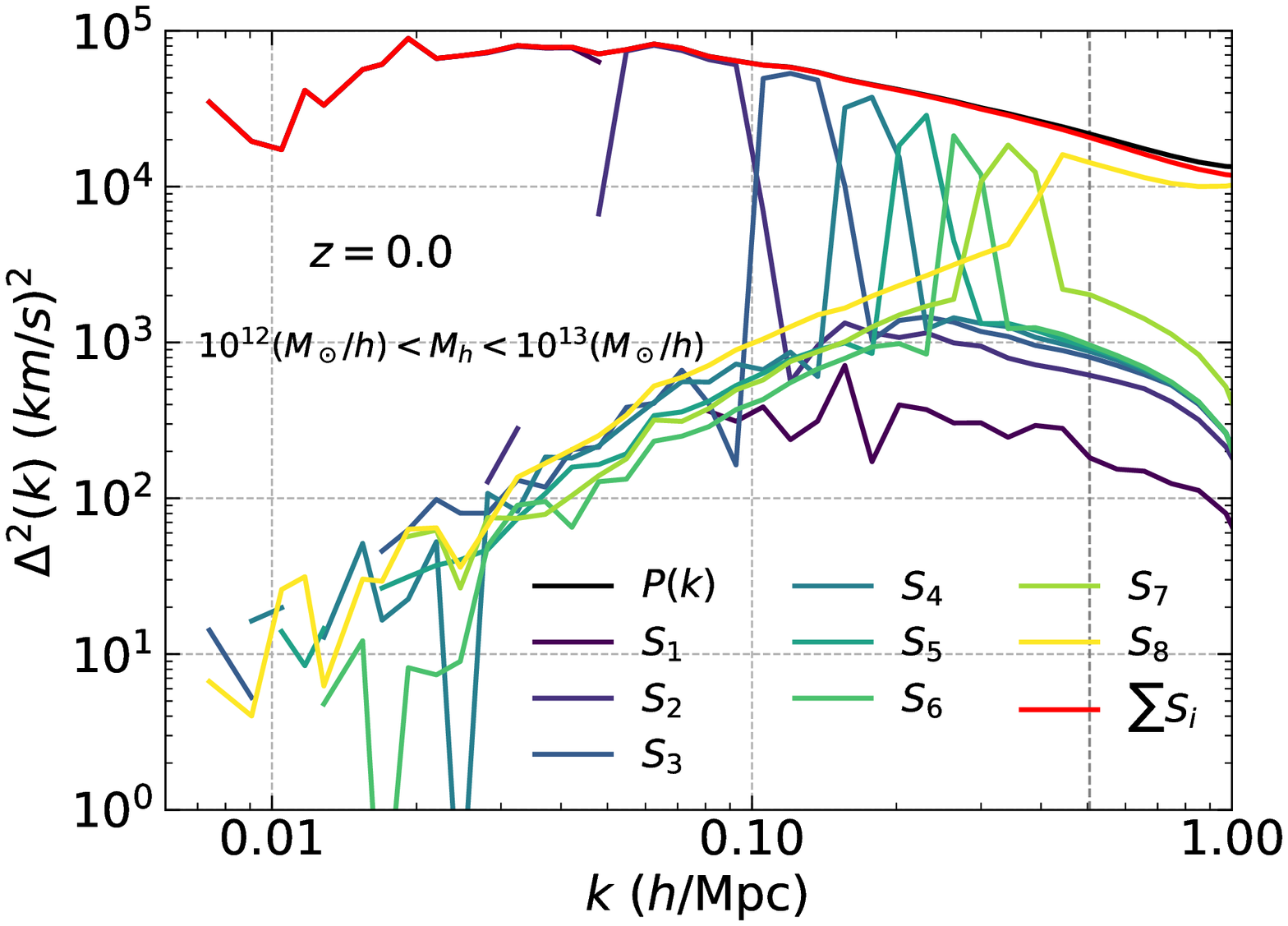}
\includegraphics[width=0.36\textwidth]{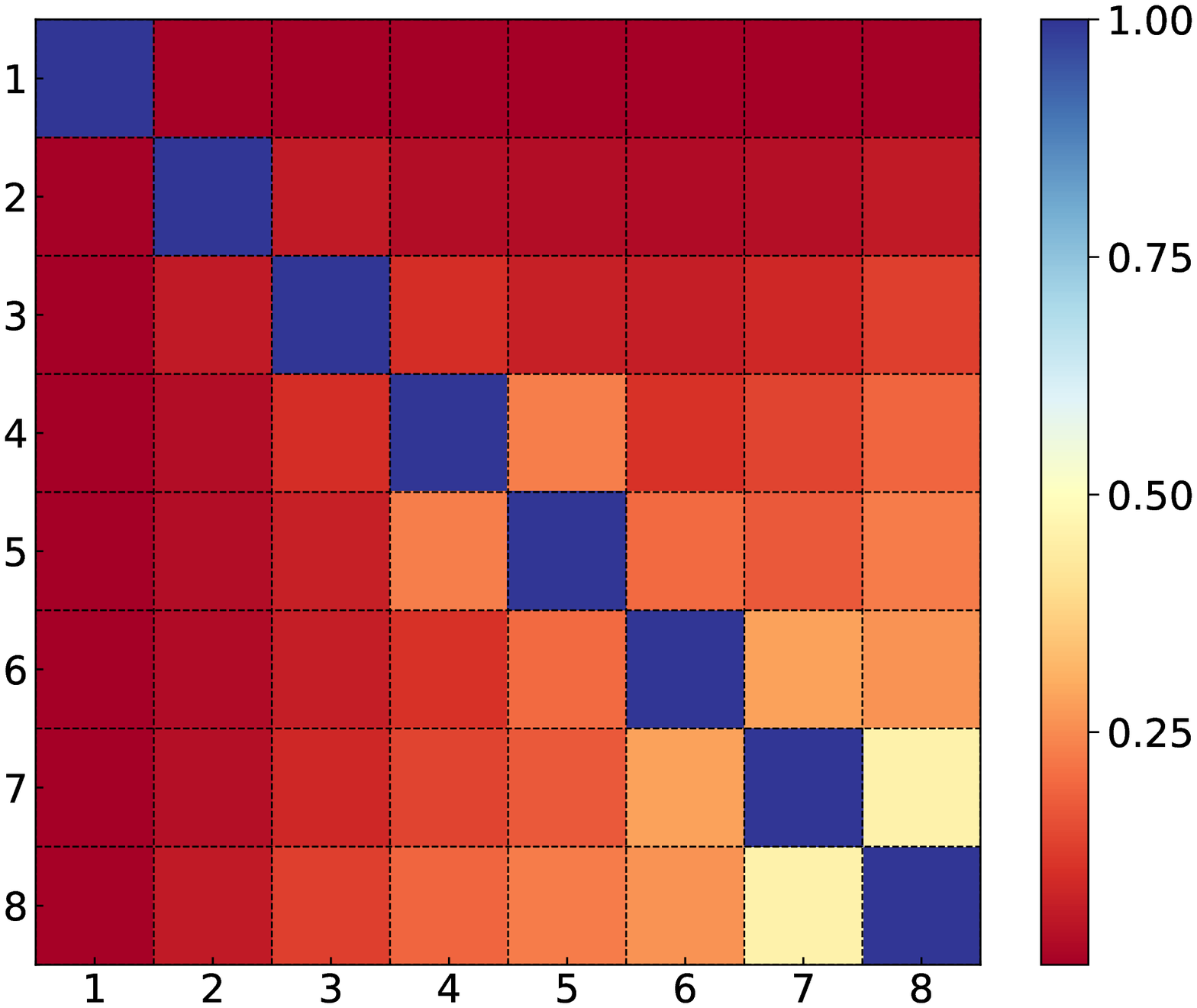}
\includegraphics[width=0.36\textwidth]{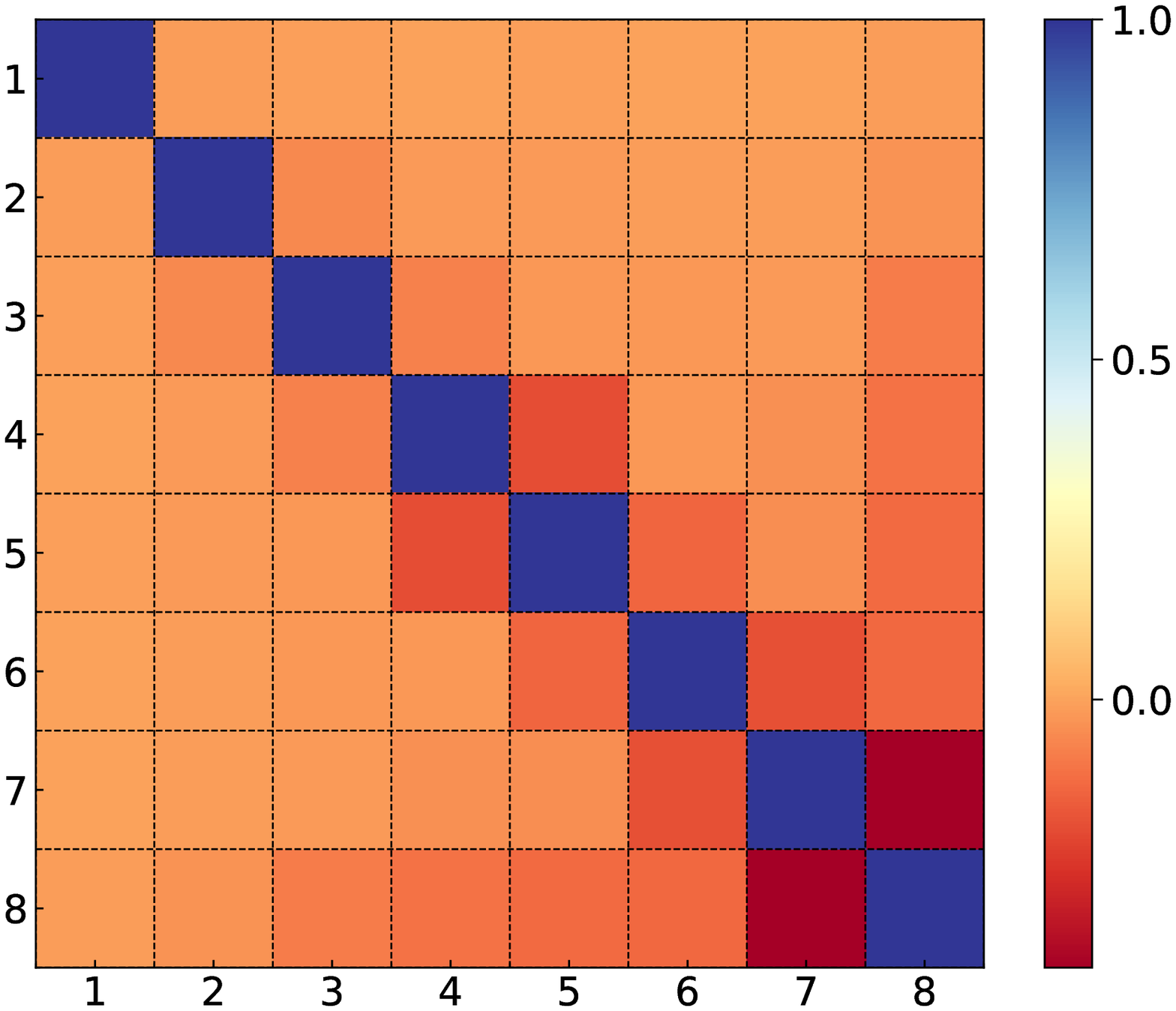}
\caption{{\bf The left panel} shows $S_\alpha(k) (\alpha=1,2,\cdots)$
  defined in Eq. \ref{eqn:S}, the sum $\sum S_\alpha$, and
  $P_{(1+\delta_h)v_hv_m}(k)$ of halos. For brevity, we only show
  the case of halos in the mass range 
  $10^{12}<M/(M_{\odot}/h)<10^{13}$ and at $z=0$.  If the halo velocity bias
  equals unity, $\sum S_\alpha=P_{(1+\delta_h)v_hv_m}(k)$. Slight
  difference between $\sum S_\alpha$ and 
  $P_{(1+\delta_h)v_hv_m}(k)$ (in particular at high $k$) implies that
  the velocity bias of corresponding halos is close to, but not
  exactly, unity. We only use the measurement at $k<0.5\hmpc$ for
  fitting $b_v$. {\bf The middle panel} shows the normalized matrix ${\bf F}_{\alpha\beta}/\sqrt{{\bf F}_{\alpha\alpha}{\bf F}_{\beta\beta}}$ of
  Eq. \ref{eqn:F}. Due
  to overlap of $S_{\alpha}$ and $S_{\beta\neq \alpha}$ in the $k$
  space, $F_{\alpha\beta}$ has non-vanishing off-diagonal
  elements. {\bf The right panel} shows the normalized error matrix ${\bf E}_{\alpha\beta}/\sqrt{{\bf E}_{\alpha\alpha}{\bf E}_{\beta\beta}}$ in
  Eq. \ref{eqn:E}.  Overlaps
  between pairs of $S_{\alpha\neq \beta}$ in the $k$ space (the left panel) cause
  ${\bf F}_{\alpha\neq \beta}\neq 0$ (the middle panel), which leads to
  ${\bf E}_{\alpha\neq \beta}\neq 0$ (the right panel). This results into correlated error in the 
  determined $b_{{1,2,\cdots}}$. It is also partly resonsible for the
  increasing statistical error in the determined $b_v$ with increasing
  $k$. \label{fig:S}}
\end{figure}

We  bin the unknown $b_v(k)$ into a
number of $k$ bins, each with central value $k_\alpha$ and bin width
$\Delta k_\alpha$ ($\alpha=1,2,\cdots$). $b_v(k)=\sum_\alpha b_\alpha
W_{\alpha}(k)$. $W_{\alpha}(k)=1$ if $k_\alpha-\Delta k_\alpha/2<k\leq
k_\alpha+\Delta k_\alpha/2$, and zero otherwise. $b_\alpha$ is the
averaged value of $b_v$ in the range $k_\alpha-\Delta k_\alpha/2<k\leq
k_\alpha+\Delta k_\alpha/2$. The power spectrum
$B_{\delta_hv_hv_m}({\bf k})
=\sum_\alpha b_\alpha B_\alpha({\bf k})$. Here $B_\alpha({\bf k})$ is
$B_{\delta_hv_hv_m}({\bf k})$  in which we replace ${\bf 
  v}_h({\bf k}^{'})$ with
${\bf v}_m({\bf k}^{'})W_\alpha(k^{'})$.  The calculation of $B_\alpha({\bf k})$ is done with
several FFTs. First we obtain ${\bf v}_m({\bf k})$ from the simulated
${\bf v}_m({\bf x})$. We then inverse Fourier transform ${\bf
  v}_m({\bf k})W_\alpha(k)$ and denote it as ${\bf
  v}_{\alpha}({\bf x})$. Finally we Fourier transform $\delta_h({\bf
  x}){\bf v}_\alpha({\bf x})$, multiply it by ${\bf v}^{*}_m({\bf k})$, 
and obtain $B_\alpha({\bf k})$. The estimated/modelled power spectrum
is then
\ba
\label{eqn:S}
\hat{P}_{(1+\delta_h)v_hv_m}({\bf k})&=&\sum_\alpha b_\alpha
  \left(W_\alpha(k)P_{v_mv_m}({\bf k})+B_\alpha({\bf k})\right) =
  \sum_\alpha b_\alpha S_\alpha({\bf k})\ ,\ S_\alpha({\bf k})\equiv
  W_\alpha(k)P_{v_mv_m}({\bf k})+B_\alpha({\bf k})\ .
\ea
Fig. \ref{fig:S} shows $S_\alpha$ for halos in the range
$10^{12}<M/(M_\odot/h)<10^{13}$. For small $\alpha$ (small
$k_\alpha$), $S_\alpha$ is dominated by the first term and is very
close to a step function in the $k$ space. But when $k$ increases, the
contribution from $B_\alpha$ becomes non-negligible. Tails beyond the
$k$ range $[k_\alpha-\Delta k_\alpha/2,k_\alpha+\Delta k_\alpha/2]$
develop.

Both $P_{v_mv_m}({\bf k})$ and $B_\alpha({\bf k})$ are measured
from the same simulation, and the only set of unknown parameters are
$b_\alpha$. Both sides are drawn from the same simulation, therefore
the determined $b_v$ will be essentially free of cosmic variance.  The only relevant statistical error in determining $b_v$ then arises from shot noise in the halo
distribution. This allows us to write down the likelihood function
$\mathcal{L}\propto \exp(-\chi^2/2)$
straightforwardly, with 
\ba
\chi^2=\sum_{\bf k} \frac{(P_{(1+\delta_h)v_hv_m} ({\bf k})-\hat{P}_{(1+\delta_h)v_hv_m}({\bf k}))^2}{\sigma^2_{\bf k}}\ .
\ea
$\sigma_{\bf k}$ is the r.m.s. fluctuation of the data
$P_{(1+\delta_h)v_hv_m} ({\bf k})$ caused by the finite number of
halos. We estimate it by splitting halos in a given mass bin into $8$ non-overlapping
sub-samples by randomly selecting among these halos.  We measure the
dispersion between these sub-samples, divide it by $\sqrt{8}$ and
obtain $\sigma_{\bf k}$.  Due to the shot noise origin,  the
errors are uncorrelated over different ${\bf k}$. 

Since $\hat{P}_{(1+\delta_h)v_hv_m}$ is a linear combination of
$b_\alpha$ (Eq. \ref{eqn:S}), the likelihood function $\mathcal{L}$ of
$b_\alpha$ is a multivariate Gaussian function.  The
best-fit of $b_\alpha$ is given by the following linear equations, 
\ba
\label{eqn:bestfit}
\frac{\partial \chi^2}{\partial b_\alpha}=0\ \Rightarrow \ 
\sum_\beta b_\beta \left[\sum_{\bf k} \frac{S_\alpha({\bf k}) S_\beta({\bf k})}{\sigma_{\bf
    k}^2}\right]=\sum_{\bf k} \frac{P_{(1+\delta_h)v_hv_m}({\bf k})S_\alpha({\bf k})}{\sigma_{\bf k}^2}\ .
\ea
The solution (best fit value)  is
\ba
\label{eqn:F}
{\bf b}={\bf F}^{-1}{\bf D}\ ,\ {\rm with}\ {\bf F}_{\alpha\beta}\equiv \sum_{\bf k} \frac{S_\alpha({\bf k})S_\beta({\bf
    k})}{\sigma_{\bf k}^2}\ \& \ {\bf D}_\alpha\equiv \sum_{\bf k}
\frac{P_{(1+\delta_h)v_hv_m}({\bf k})S_\alpha({\bf k})}{\sigma_{\bf k}^2}\ .
\ea
The error covariance matrix
${\bf E}_{\alpha\beta}\equiv \langle \delta b_\alpha\delta b_\beta\rangle$ is  given by
\ba
\label{eqn:E}
{\bf E}_{\alpha\beta}=(F^{-1})_{\alpha\beta}\ \ .
\ea
The matrix ${\bf F}$ and ${\bf E}$ are shown in Fig. \ref{fig:S}. Due
to overlap of $S_{\alpha}$ and $S_\beta$ ($\beta\neq \alpha$) in the
$k$ space, ${\bf F}$ has non-diagonal elements. They result in
correlated error in the determined $b_\alpha$, quantified by
${\bf E}_{\alpha\neq \beta}\neq 0$. The correlation is stronger for pairs of
larger $k_\alpha$ and $k_\beta$. 
\label{eqn:solve}

\bibliographystyle{apj}
\bibliography{mybib}

\end{document}